\input{aipcheck}

\documentclass[
    ,final            
  ]
  {aipproc}

\usepackage[dvips]{color}
\usepackage{amsmath}
\usepackage{amsfonts}
\usepackage{amssymb}

\usepackage{graphics}
\usepackage{amsmath}
\usepackage{amsfonts}
\usepackage{amssymb}
\usepackage{graphicx}
\usepackage{epsfig}

\def\ba{\begin{eqnarray}}
\def\ea{\end{eqnarray}}
\def\be{\begin{equation}}
\def\ee{\end{equation}}

\def\d{\mathrm{d}}

\def\mn{_{\mu \nu}}

\def\({\left(}
\def\){\right)}

\def\ie{{\it i.e. }}

\definecolor{dunkelblau}{rgb}{0, 0, .7}
\definecolor{dunkelmagenta}{rgb}{1, 0.2, 1}
\definecolor{dunkelmagenta}{rgb}{0.95, 0.2, 0.8}
\definecolor{dunkelmagenta}{rgb}{1, 0, 0}
\definecolor{cyan}{rgb}{0,0,0.7}
\definecolor{cyand}{rgb}{0,0,0.7}

\layoutstyle{6x9}

\begin{document}

\title{Classical Renormalization of Codimension-two Brane Couplings}

\classification{11.10.Gh, 11.10.Kk, 11.25.-w}
\keywords      {large extra dimensions, six-dimensional gravity, brane couplings, renormalization}

\author{Claudia de Rham}{
address={Dept. of Physics \& Astronomy, McMaster University, Hamilton ON, Canada\\
Perimeter Institute for Theoretical Physics,  Waterloo, ON, Canada}
}
%
%

\begin{abstract}
The curvature on codimension-two and higher branes is not
regular for arbitrary matter sources. Nevertheless, the low-energy theory
for an observer on such a brane should be well-defined and
independent to any regularization procedure. This is achieved via
appropriate {\it classical} renormalization of the brane
couplings, and leads to a natural hierarchy between standard model
couplings and couplings to gravity.
\end{abstract}

\maketitle

\subsection{General matter sources on codimension-two branes}


Codimension-two braneworld models have recently enjoyed a increased
interest in their potential resolution of the cosmological constant
problem, \cite{Rubakov:1983bz}.
However, understanding the cosmology on such
objects, requires the introduction of matter sources other than pure
tension, which
has been shown to lead to metric divergences on the defect, \cite{Geroch:1987qn}.
%
%
As a consequence, it is necessary to regularize codimension-two
branes before drawing any physical conclusion.
A priori this is not a problem since any codimension-two object arises from an underlying theory which
naturally provides a regularization scheme. However, we still expect from field
theory that the low-energy theory on the brane is
independent from such a regularization. Or in other words, at low-energy, one
should be able to integrate out the fields responsible for the
object regularization, and thus end up with a low-energy physics
independent to the high-energy regularization scale.
To this end, the same technique as that usually
employed in field theory should be followed and the couplings of the
theory should be renormalized, \cite{Goldberger:2001tn}.

\subsection{Origin of the Problem}

To understand the origin of the problem, let us consider a free
massless scalar field $\phi$ (symbolizing the graviton) living in a flat
six-dimensional space-time
\begin{eqnarray}
\d s^2=\eta\mn \d x^\mu \d x^\nu+\d r^2 +r^2 \d \theta^2,
\end{eqnarray}
with $0\le\theta< 2\pi \alpha$, and where $(1-\alpha)$
is the deficit angle ($\alpha \le 1$). Its propagator $D(x^a,x'^a)$,
satisfying $r \Box_x^{(6d)} D(x^a,x'^a)=i \delta^{(6)}(a^a-x'^a)$ is
finite as long as one of the end point is taken away from $r=0$,
but it contains a logarithmic singularity when
trying to evaluate both points on the tip of the cone.
Introducing a momentum cutoff scale $\Lambda$ in the evaluation of
the propagator, one has (see ref.~\cite{Goldberger:2001tn}),
\ba
\label{b-b prop 2}
D_k(0,0)=-\int_0^{\Lambda}\frac{\d q q}{2 \pi \alpha}
\frac{i}{k^2+q^2}=\frac{-i}{2\pi \alpha}\log
\frac{\Lambda}{k}\,,
\ea
where $k^\mu$ is the four-dimensional momentum.
In principle, this should not be a problem, unless a physical source
is introduced at $r=0$. In that case,
gravity diverges on the brane.
We can therefore wonder what
observers on such a codimension-two brane located at $r=0$ and coupled to
gravity will feel.

\subsection{Observer on the brane}
To mimic the effect on observers on the brane, we consider a second
scalar field $\chi$, that we call the Standard Model (SM) field, localized on the tip of the cone and coupled to
``gravity" (the field $\phi$):
\ba
\label{FreeAction2}
S=-\int \d ^4 x\d \theta \d r\ r\(\frac12 (\partial_a
\phi)^2+\delta^2(y)\left[
\frac{1}{2} (\partial_\mu \chi)^2 +\frac{m^2}{2} \chi^2 +\lambda \chi \phi\right]
\).
\ea
In the absence of the coupling $\lambda$ between the SM field
$\chi$ and the graviton $\phi$, the SM propagator
would simply be $G^{\chi\chi}_k=-i(k^2+m^2)^{-1}$. The coupling with gravity
will however induce divergences to the SM propagator, leading to the following ``dressed"
two-point function, \cite{de Rham:2007pz}
\ba
G^{\chi\chi}_k=-\frac{i}{(k^2+m^2)-i\lambda^2
 D_k(0,0)}\,.
 \ea
This clearly shows how the coupling of gravity to the SM field
introduces a singular term $D_k(0,0)$ in the expression of the SM
propagator.
As it stands, one can consider arbitrarily small
energies, the propagator will still depend on the cutoff scale
$\Lambda$, through $D_k(0,0)=-i/2\pi\alpha \log \Lambda/k$. To make
sense of the theory at low energies, $k\gg\Lambda$,
this divergence ought therefore to be absorbed in the coupling
constants. In particular the divergence disappears  if the SM mass $m$ is
renormalized in the following way
\ba
\label{RenormCouplingsm}
m^2(\mu)=m^2(\Lambda)-\frac{1}{2 \pi \alpha}\, \lambda^2(\Lambda)\log\frac{\Lambda}{\mu},\hspace{40pt}
\ea
leading to the following Renormalization Group (RG) flows
\ba
\mu \frac{\d m^2(\mu)}{\d \mu}=\frac{\lambda^2(\mu)}{2\pi
\alpha}\,,\hspace{30pt}\mu \frac{\d \lambda(\mu)}{\d \mu}=0
\ea
where $\mu$ is the physical scale. This mass renormalization is
sufficient to ensure that the propagators for both fields are
finite and independent of the cutoff scale. The Green's function of
the matter field on the brane is thus finite despite being coupled with the graviton field which
itself has not a well-defined limit on the brane.

\subsubsection{Further interactions}
To this setup, further brane-bulk interactions can also be considered, which
will in turn affect the brane observables.
In particular, restricting ourselves to the relevant and marginal operators,
the most general brane interactions are then
\ba
S&=&-\hspace{-3pt}\int \d^6x\, \Big[
\frac 12 \(\partial \phi\)^2
+\frac{\delta(r)}{2r\pi \alpha}\,\mathcal{L}_{\chi \phi}
\Big]\,,\\
\mathcal{L}_{\chi \phi}&=&\(\frac 12 \(\partial\chi\)^2+\frac 12 m^2 \chi^2+\beta_3 \chi^3+\beta_4 \chi^4\)
+\(\lambda \chi\phi+\lambda_2\chi^2 \phi\)\,.
\ea
At this point two independent kinds of divergences can be distinguished.
Interactions of the form  $\beta_n \chi^n$ with $n>2$, induce loop
corrections to the standard Green's functions which diverge in the
ultra-violet. These divergences, standard in field theory, can be
absorbed by renormalization of the mass $m$, the coupling
$\beta_4$ as well as the wave-function. However, such divergences
can be treated in a completely independent way to that arising at
the tree-level in our codimension-two scenario.
Interactions of the form $\lambda_3 \phi \chi^2$, for instance
will typically induce tree-level divergences which
can be absorbed by appropriate renormalization of the couplings.

At tree-level, the divergent parts of the brane field three and four-point
function are proportional to
\ba
\langle \chi^3\rangle_{\text{div}}&\propto& \(\beta_3
-i \lambda_3 \lambda D_{k}(0,0)\)\\
\langle \chi^4\rangle_{\text{div}}&\propto& \(\beta_4
-\frac i 2 \lambda_3^2 D_{k}(0,0)\)\,,
\ea
which again can be absorbed by appropriate renormalization of the
couplings $\beta_3$ and $\beta_4$:
\ba
\mu \frac{\d \beta_3(\mu)}{\d \mu}=\frac{\lambda(\mu)\lambda_3(\mu)}{2\pi
\alpha}\,,\hspace{30pt}\mu \frac{\d \beta_4(\mu)}{\d \mu}=\frac{\lambda_3(\mu)^2}{4\pi
\alpha}\,,
\ea
while the coupling $\lambda_3$ remains constant $\partial_\mu
\lambda_3=0$, similarly as $\lambda$. Furthermore, these coupling
flows is actually sufficient to ensure that every $n$-point function
(with $n\ge2$) is finite as the codimension-two cutoff scale
$\Lambda$ is sent to infinity, \cite{de Rham:2007pz}. With respect
to this divergence, the theory is thus completely renormalizable.
We have therefore achieved to derive a low-energy effective field
theory on the codimension-two brane, independent to any regularization procedure.

\subsubsection{Hierarchy problem}
Already at the level of this simple toy-model, one can distinguish between two
different kinds of couplings:
\begin{itemize}
\item Bulk-Brane couplings $\lambda_B$, ($\lambda_B=\lambda, \lambda_3,\cdots$),  symbolizing interaction with gravity, that do not
flow in this specific example, $\partial_\mu \lambda=\partial_\mu
\lambda_3=0$,
\item Pure brane couplings $\beta_b$, ($\beta_b=m, \beta_3,\beta_4,\cdots$), symbolizing interactions from the SM, which typically flow as
$\mu \partial_\mu \beta \propto \lambda \lambda_B$.
\end{itemize}
Couplings from the SM thus behave very differently from couplings with bulk fields (\ie
gravity), leading to the important conclusion that interactions with bulk fields
are naturally suppressed compared to interactions of brane
fields between themselves! Although this suppression is not sufficient to solve the hierarchy problem by
itself, it can act constructively to the ADD scenario, \cite{Arkani-Hamed:1998rs}.
Moreover, in higher-codimension scenarios, the suppression could be
much more significant, since the divergence in this case is that of
a power law rather than logarithmic. This could potentially lead to a potential
hierarchy between SM forces and gravity in braneworld scenarios with
codimension greater than two.

\subsection{Discussion}
So far, distributional sources on codimension-two branes only made sense as regularized
objects, since bulk fields typically diverge logarithmically when evaluated on the
brane. We have presented here how classical regularization of brane
couplings lead to well-defined observables on the brane, and this
independently to any regularization procedure. This
provides a well defined low-energy effective theory on the brane
which resulting RG flows can offer interesting physical
implications. A hierarchy between the SM forces and gravity appears for instance as a natural
consequence, and could be amplified in higher-codimension setups.


\begin{theacknowledgments}
Research at McMaster is supported by the Natural
Sciences and Engineering Research Council of Canada.
Research at Perimeter Institute for Theoretical Physics is supported
in part by the Government of Canada through NSERC and by the
Province of Ontario through MRI.
\end{theacknowledgments}



\bibliographystyle{aipproc}   


\IfFileExists{\jobname.bbl}{}
 {\typeout{}
  \typeout{******************************************}
  \typeout{** Please run "bibtex \jobname" to optain}
  \typeout{** the bibliography and then re-run LaTeX}
  \typeout{** twice to fix the references!}
  \typeout{******************************************}
  \typeout{}
 }

\end{document}